# Estimating SARS-CoV-2 transmission in educational settings: a retrospective cohort study

Mattia Manica [1,2,§], Piero Poletti [1,2,§], Silvia Deandrea [3,§], Giansanto Mosconi [3,4], Cinzia Ancarani [3], Silvia Lodola [3], Giorgio Guzzetta [1,2], Valeria d'Andrea [1], Valentina Marziano [1], Agnese Zardini [1], Filippo Trentini [1,5], Anna Odone [4], Marcello Tirani [6,#], Marco Ajelli [7,#], Stefano Merler [1,2,#,*]

[§] equally contributed
[#] joint senior authors
[*] corresponding author: merler@fbk.eu

[1] Center for Health Emergencies, Bruno Kessler Foundation, Trento, Italy
[2] Epilab-JRU, FEM-FBK Joint Research Unit, Trento, Italy
[3] Prevention Department, Agency for Health Protection, Pavia, Italy
[4] Department of Public Health, Experimental and Forensic Medicine, University of Pavia, Pavia, Italy
[5] Dondena Centre for Research on Social Dynamics and Public Policy, and CovidCrisisLab, Bocconi University, Milan, Italy
[6] Directorate General for Health, Lombardy Region, Milan, Italy
[7] Laboratory for Computational Epidemiology and Public Health, Department of Epidemiology and Biostatistics, Indiana University School of Public Health, Bloomington, IN, USA

## Abstract

**Background** School closures and distance learning have been extensively applied to control SARS-CoV-2 transmission. Despite evidence of viral circulation in schools, the contribution of students and of in-person schooling to the transmission remains poorly quantified.

**Methods** We analyze 976 exposure events, involving 460 positive individuals, as identified in early 2021 by routine surveillance and through an extensive screening conducted on students, school personnel, and their household members during an outbreak in a small municipality of Italy.

**Results** From the analysis of potential transmission chains, we estimated that, on average, 55.1%, 17.3% and 27.6% infection episodes were linked to household, school, and community contacts, respectively. Clusters originated from students or school personnel showed a larger average cluster size (3.32 vs 1.15), a larger average number of generations in the transmission chain (1.56 vs 1.17) and a larger set of associated close contacts (11.3 vs 3.15, on average). We found substantial transmission heterogeneities, with 20% positive individuals seeding 75-80 of all transmission. A higher proportion of infected individuals causing onward transmission was found among students (48.8% vs 29.9%, on average), who also caused a markedly higher number of secondary cases (mean: 1.3 vs 0.5).

**Conclusions** Uncontrolled transmission at school could disrupt the regular conduct of teaching activities, likely seeding the transmission into other settings, and increasing the burden on contact-tracing operations.

## Background

School closures and the replacement of in-person school attendance with distance learning were extensively implemented during the first two years of the COVID-19 pandemic. Such policy has been associated with a heated public debate due to its impact on the quality of students' education, the costs and resources required to provide safe educational environments, and the well-being of children and their parents (European Centre for Disease Prevention and Control, 2021; Flasche and Edmunds, 2021). Compared to adults, children are less likely to develop symptoms and thus of being reported to surveillance systems (Poletti et al. 2020; Ladhani and the sKIDs Investigation Team, 2021; Flasche and Edmunds, 2021). However, SARS-CoV-2 transmissibility is not associated with age (Hu et al., 2021) and evidence of viral circulation in schools has been repeatedly found (Goldstein et al., 2021; Flasche and Edmunds, 2021; Larosa et al., 2020; Theuring et al., 2021, Meuris et al., 2021). The quantification of the contribution of school to the overall SARS-CoV-2 transmission remains elusive. Most studies are based on the analysis of epidemiological trends observed after schools' reopening, and therefore at risk of confounding and collinearity from the simultaneous release of other non-pharmacological interventions (Walsh et al., 2021). School-related disease incidence was found to increase with the proportion of students receiving in-person education (Doyle et al., 2021), while, at the same time, infection prevalence among students and teachers was associated with COVID-19 incidence in the community (Ismail et al., 2021; Tupper and Colijn, 2021).

In this study, we analyze 460 SARS-CoV-2 positive individuals and 976 contacts identified by routine surveillance and through an extensive screening of the student population and of their households during an outbreak in Mede, Italy, in early 2021. The analysis of the reconstructed transmission chains allowed us to provide quantitative estimates of the role of students in the spread of the infection, considering potential heterogeneities in the risk of infection and onward transmission after the exposure to SARS-CoV-2 in households, schools, and the community.

## Methods

### Study population

In February 2021, a rapid upsurge of symptomatic cases was detected in Mede, a small municipality of the Lombardy region in Italy (6,326 inhabitants). A non-negligible set of positive cases were identified among students, raising concern about widespread circulation of SARS-CoV-2 in schools. The outbreak involved a crèche, a kindergarten, and the local main school, which consists of a primary (259 pupils) and a middle school (155 students). On February 5, 2021, all students of three classes (two from the kindergarten and one from the primary school) were isolated at home according to protocols implemented in the country to prevent school outbreaks (Ministero della Salute, 2020). The progressive increase of cases determined the closure of the kindergarten on February 7, the closure of the crèche, the primary and the secondary school on February 16, and the application of the highest level of restrictions to the whole municipality on February 17. Between February 17 and March 23, local health authorities carried out a free screening campaign based on PCR tests and involving all individuals connected to the schools (i.e., students/school personnel and their household members). Information about the household of screened individuals, the class and the school attended by tested students was recorded. The screening was followed by the tracing and testing of contacts of identified SARS-CoV-2 positive individuals. The definitions of COVID-19 cases and case contacts adopted in Italy can be found in (Poletti et al., 2021).

The analyzed sample consists of all SARS-CoV-2 positive individuals and their contacts identified between January 7 (the date of school reopening after the Christmas holidays) and March 10, 2021 (i.e., 3 weeks after the strengthening of restrictions in the whole municipality). Positive individuals were classified as symptomatic infections if they showed upper or lower respiratory tract symptoms or fever ≥37.5 °C (Poletti et al., 2021). Respiratory symptoms included dry cough, dyspnea, tachypnea, difficulty breathing, shortness of breath, sore throat, and chest pain or pressure (Poletti et al., 2021). Figure 1A shows the time series of symptomatic cases identified in Mede during the study period, along with the timeline of interventions implemented in the municipality.

### Reconstruction of the transmission chains

We defined as cluster, a set of two or more SARS-CoV-2 positive individuals having an epidemiological link, as identified during contact tracing activities. Inside each cluster, the transmission chain was inferred by using the following procedure. Potential infection episodes were identified by collating for each positive individual the set of their possible infectors, based on close contacts reported by ascertained infections during epidemiolocal investigations. To deal with multiple exposures, we randomly sampled the potential infector among the pool of positive contacts. Specifically, infectors were assigned with equal prior probability between possible links while checking for consistence in the resulting transmission chain (i.e., rejecting circular transmission within the analyzed clusters, see Figure 1B). The entire procedure is repeated 1,000 times to obtain different instances of consistent transmission chains. Each simulated instance was analyzed in terms of i) the size of identified clusters of infection, ii) the length of different transmission chains, defined as the number of generations within clusters, iii) the number and age of secondary cases caused across different settings by infectors of different ages, iv) the risk of resulting positive after an exposure event. To define the likely source of infection, we assumed that exposures occurring between cohabiting individuals took place in their household. Exposures recorded between individuals attending the same school (either as students or school personnel) but not sharing the same household were considered as school exposures. Exposures outside the household and school were classified as occurred in the community. Positive cases without a history of exposure to SARS-CoV-2 were assumed to be associated to an unknown source of infection in the community. To check for heterogeneity in transmission, we fit a negative binomial distribution to the offspring distribution estimated by calculating the number of secondary cases caused by each positive individual. Obtained results are presented in terms of average values and 95% prediction intervals (PI) computed over the 1,000 instances of the transmission chains.

## Results

### *Descriptive statistics*

We collated a dataset of 822 individuals tested for SARS-CoV-2 infection between January 7 and March 10, 2021, representing 13% of the residents of the focus municipality (Mede). The median age of the tested individuals was 36 years (IQR: 11-53, ranging from 1 month to 98 years); 52.3% individuals were female. Out of the 822 tested individuals, 460 resulted PCR positive for SARS-CoV-2 infection, including 237 (51.5%) showing symptoms while 183 (39.8%) were asymptomatic (for 40 positive individuals, 8.7%, this information was not reported). The median age of symptomatic and asymptomatic cases was 43 years (IQR: 28-56) and 31 years (IQR: 10-49), respectively. Among the ascertained infections, 311 were identified through standard surveillance and 149 through the scholastic investigation. The positivity ratio in these two groups of tested individuals was 66.9% (311/465) and 41.7% (149/357), respectively. The corresponding symptomatic ratio was 57.9% (180/311) and 38.3% (57/149). Overall, 203 (24.7%) tested individuals were students and 13 (1.6%) were school personnel. Out of the 460 ascertained infections, 82 (17.8%) were students and 8 (1.7%) were school personnel. A detailed description of the analyzed sample is reported in Table 1.

The average number of close contacts reported by positive individuals with household members was 1.1 (IQR: 0-2); 0.6 (IQR: 0-1) contacts per person were identified in the community. Positive students reported an additional 3.2 (IQR: 0-2) contacts with schoolmates or school personnel, resulting in a higher average number of close contacts experienced overall (5.0 vs 1.5 for non-students; Wald-test p-value < 0.0001). The average number of contacts that an individual of a given age has with individuals in other age groups can be visualized in the form of a matrix (Figure 1C). The obtained contact matrix shows that the highest contact rate was reported by individuals aged 10-20 years with individuals in the same age group.

We identified 976 potential exposure events: 432 consisted of either single or multiple negative exposure events, 218 led to the identification of a clear infection episode, 326 were associated with positive individuals reporting contacts with multiple positive individuals. 250 exposures were excluded because of incomplete information on the involved peers. Information about the age, sex, household, and school of cases and their close contact was instead available for 726 exposures, involving 221 potential infectors and 627 contacts. Of these contacts, 261 tested positive and 366 tested negative.

*SARS-CoV-2 transmission patterns*

From the analysis of potential transmission chains (Figure 2), we identified 144 (95%PI: 140-148) clusters of infections. The identified clusters were associated with an average number of 4.49 (95%PI: 4.35-5.64) close contacts up to a maximum of 81.1 (95%PI: 76-87). The average cluster size (defined as the mean number of infection episodes identified per cluster) was 1.49 (95%PI:1.42-1.57) up to a maximum size of 28.3 (95%PI: 25-32). The mean number of generations observed per cluster was 1.23 (95%PI: 1.20-1.27).

Clusters originating from students or school personnel consisted of a larger number of generations in the transmission chain (1.56, 95%PI: 1.42-1.70 vs 1.17, 95%PI: 1.13-1.21). The average number of generations in clusters including at least one positive student was 1.85 (95%PI: 1.72-2.) compared to 1.09 (95%PI: 1.06-1.11) generations in clusters with no infected students. Most of the clusters (121, 95%PI: 117-125) originated from individuals not related with the scholastic setting; 23 (95%PI: 21-25) clusters originated from students or school personnel and showed a larger average cluster size (3.32, 95%PI: 2.79-3.83 vs 1.15, 95%PI: 1.06-1.24) and a larger set of associated close contacts (11.3, 95%PI: 10.3-12.9 vs 3.15, 95%PI: 2.95-3.35).

We estimated that 118 (95%PI: 110-125) infection episodes were linked to a household contact, out of the 291 (95%PI: 283-298) estimated household exposures; 37 (95%PI: 36-40) infections were linked to a transmission in school (out of the 170, 95%PI: 168-172, estimated exposures); 59 (95%PI: 55-64) infections occurred in the community (185, 95%PI: 181-190, estimated exposures. Accordingly, infection episodes represented 40.5% (95%PI: 38.9-41.9%), 22.2% (95%PI: 21.4-23.3%), and 32% (95%PI: 30.4-33.7%) of all estimated exposures occurred in the household, school, and community, respectively.

We found that 154 (95%PI: 146-161) positive individuals (42.9%, 95%PI: 41.7-44.1% of the analyzed potential infectors) did not cause any secondary infection. A higher proportion of individuals causing onward transmission was found among positive students (48.8% vs 29.9%, on average). The average number of secondary infections caused by a positive individual was estimated to be 0.6 (95%PI: 0.59-0.61) – we stress that the average number of secondary infections must be lower than 1 in any contained outbreak (Lloyd-Smith et al., 2005). We estimated the distribution of the number of secondary infections to follow a negative binomial distribution with overdispersion (shape parameter) 0.53 (95%PI: 0.47-0.61), implying that 20% of infectors were responsible for 75-80% of all secondary cases (Figure 3A). A similar heterogeneity in the transmission was found among students (overdispersion: 0.53, 95%PI: 0.45-0.62). However, positive students caused on average a markedly higher number of secondary cases (mean: 1.26, 95%PI: 1.18-1.33). No relevant differences were found in the number of secondary infections caused by school personnel and by individuals unrelated with the school setting (0.37 vs 0.43, on average). The average number of secondary cases caused by any positive individual at home and in the community was 0.33 (95%PI: 0.31-0.34) and 0.17 (95%PI: 0.15-0.18), respectively. Positive students caused an additional 0.5 (95%PI: 0.47-0.52) cases among school-related contacts (schoolmates or school personnel).

Based on the identified infection episodes we reconstructed an age-specific matrix representing the average number of infections caused in each age group by a positive case, stratified by the age of the infector (Figure 3B). The highest transmission intensity was found from young children to adults, and between children of similar age (possibly reflecting contacts between siblings or schoolmates). The ratio between the number of secondary cases and the number of reported exposures was markedly lower for interactions occurred at school (29.1% vs 43.8% in household, and 34.7% in the community).

## Discussion and conclusions

In this work, we analyzed a scholastic outbreak in an Italian municipality. The collected records included PCR positive individuals and their close contacts identified by routine surveillance and through an extensive screening conducted on students, school personnel, and their household members. By reconstructing the transmission chains, we investigated the role of students in SARS-CoV-2 transmission.

Despite protocols in place to curb SARS-CoV-2 transmission during in-person school attendance, we found that younger age groups were deeply involved in the spread of the infection. The average number of secondary cases caused by an infected student was significantly larger as compared to other individuals (1.3 vs 0.5). This result well compares with evidence from France suggesting that, in spring 2021, the school-specific reproductive number was significantly higher than that estimated for the community (Colosi et al., 2021). About 10% of the identified infection episodes occurred because of interactions between schoolmates. The transmission between schoolmates was associated with longer transmission chains, a larger number of individuals was exposed to the infection (on average 5.0 contacts were named by positive students as compared to 1.5 for other individuals), and a larger number of infections was estimated for cluster originated from school related exposures. In sum, the circulation of SARS-CoV-2 in the scholastic population entails a high risk of importation of the infection into a large set of households from where it can reach higher-risk segments of the population (Hu et al., 2021). In addition, uncontrolled transmission in the student population could disrupt the regular conduct of teaching activities and leading to a harsh burden for contact-tracing operations at the same time.

In line with previous studies (Adam et al., 2020; Colosi et al., 2021; Lemieux et al., 2021; Sun et al., 2021), the estimated distribution of secondary infections indicates a substantial transmission heterogeneity, with 20% of positive individuals causing 75-80% of all transmission events. This heterogeneity suggests that control programs targeting contexts responsible for most of the transmission could be effective in limiting the spread of the infection, including educational settings (Adam et al., 2020; Colosi et al., 2021; Woolhouse et al., 1997).

A key limitation of our study is that we were not able to collect the time of exposure(s) for positive individuals. On the one hand, this prevented us to apply standard Bayesian approaches to reconstructing the occurred transmission chains (Guzzetta et al., 2020; Meuris et al., 2021). On the other hand, when an infection episode was identified between individuals involved in the scholastic settings, we cannot exclude that the transmission occurred outside school. In our analysis, potential infection episodes were identified by sampling from competitive exposures of positive individuals across different settings, based on the set of close contacts reported by each positive individual. As testing and screening efforts across different settings were unbalanced, biases in the attribution of the infector for individuals with an unclear source of infection cannot be excluded. As for other epidemiological investigations, it is likely that some contacts were not identified or remained untested. Finally, data presented here refer to the first months of 2021, when the Alpha variant was emerging in Italy and the vaccination rate was sharply increasing. However, the impact of vaccination during the study period should be negligible (less than 16% of the population was vaccinated with 1 dose by the end of the study period (Istituto Superiore di Sanità, 2021)). Finally, the analyzed data do not provide sufficient granularity to provide estimates of school transmission vs. transmission between schoolmates, which may potentially occur in other social settings.

The contribution of students to SARS-CoV-2 transmission as well as the high prevalence in the student population we have found through an extensive screening of the student population hint at the difficulties in tracking asymptomatic infections and the challenges implementing reactive class/school closures to interrupt SARS-CoV-2 transmission (Liu et al., 2022; Colosi et al., 2021; Torneri et al., 2021, Tupper and Colijn, 2021). The surge of COVID-19 cases observed throughout Europe in mid-March 2022 and the progressive shift of the age of cases towards younger ages highlight the need of closely monitoring the epidemiological situation in schools and the community at large.

## Acknowledgments

The authors wish to thank Mara Azzi, Lorella Cecconami, Santino Silva and Stefano Boni, who were in charge of the strategic management of the Pavia Healthcare Agency and whose support made this project possible. The authors also thank the Contact Tracing Team (Alessandro Carlesso, Claudio D'Amico, Roberto Madonna, Erika Lucente, Antonietta Sabetta, Silvia Serangeli, Haydee Suarez) and the Swab Team (Sergio Edo, Ivana Menardo, Mattia Morelli, Samar Sozzi) and the General Practitioners from Mede municipality, who performed the surveillance and control activities described in this study.

## Conflict of interest

MA has received research funding from Seqirus. The funding is not related to COVID-19. All other authors declare no competing interest.

## Authors' contributions:

PP, MT, MA and SM conceived the study. PP and MM wrote the first draft of the manuscript. MM wrote the code and performed the analyses. SD, GM, CA, SL, MT collected data. MT, SD verified all data. GG, VdA, VM, AZ, FT contributed to data interpretation. PP, SD, MT, MA and SM supervised the study. All authors read, reviewed, and and approved the final version of the manuscript.

## Funding:

European Commission Horizon 2020 (Verdi - Grant agreement ID: 694160)

## Ethics approval and consent to participate:

Data collection and analysis were part of outbreak investigations conducted during a public health emergency. The processing of COVID-19 data is necessary for reasons of public interest in the area of public health, such as protecting against serious cross-border threats to health or ensuring high standards of quality and safety of health care; therefore, this study was exempted from institutional review board approval (Regulation EU 2016/679 GDPR). The school setting screening was performed upon informed consent of participants, pursuant the directive issued by Lombardy Region (Prot. G1.2021.0013306 02/03/2021).

## Availability of data and material:

Aggregate data analysed during this study will be included in this published article as supplementary information files.

## Consent for publication

Not appplicable

# Figures and Tables

**Table 1** Description of the analyzed sample.

| | | | Tested | SARS-CoV-2 positive (%) | Symptomatic cases (%) |
|---|---|---|---|---|---|
| ***Overall*** | **Overall** | | 822 | 460 (56%) | 237 (51.5%) |
| | Age class (years) | 0 - 5 | 55 | 17 (30.9%) | 7 (41.2%) |
| | | 6 - 10 | 84 | 42 (50%) | 9 (21.4%) |
| | | 11 - 20 | 153 | 68 (44.4%) | 30 (44.1%) |
| | | 21 - 35 | 104 | 69 (66.3%) | 34 (49.3%) |
| | | 36 - 50 | 191 | 120 (62.8%) | 75 (62.5%) |
| | | 51 - 65 | 108 | 63 (58.3%) | 42 (66.7%) |
| | | above 66 | 127 | 81 (63.8%) | 40 (49.4%) |
| | Sex | Female | 430 | 257 (59.8%) | 140 (54.5%) |
| | | Male | 392 | 201 (51.3%) | 97 (48.3%) |
| ***Scholastic screening*** | **Overall** | | 357 | 149 (41.7%) | 57 (38.3%) |
| | Age class (years) | 0 - 5 | 38 | 10 (26.3%) | 5 (50%) |
| | | 6 - 10 | 57 | 25 (43.9%) | 4 (16%) |
| | | 11 - 20 | 90 | 30 (33.3%) | 13 (43.3%) |
| | | 21 - 35 | 44 | 28 (63.6%) | 9 (32.1%) |
| | | 36 - 50 | 68 | 30 (44.1%) | 17 (56.7%) |
| | | 51 - 65 | 29 | 11 (37.9%) | 7 (63.6%) |
| | | above 66 | 31 | 15 (48.4%) | 2 (13.3%) |
| | Sex | Female | 186 | 89 (47.8%) | 36 (40.4%) |
| | | Male | 171 | 60 (35.1%) | 21 (35%) |
| ***Routine surveillance*** | **Overall** | | 465 | 311 (66.9%) | 180 (57.9%) |
| | Age class (years) | 0 - 5 | 17 | 7 (41.2%) | 2 (28.6%) |
| | | 6 - 10 | 27 | 17 (63%) | 5 (29.4%) |
| | | 11 - 20 | 63 | 38 (60.3%) | 17 (44.7%) |
| | | 21 - 35 | 60 | 41 (68.3%) | 25 (61%) |
| | | 36 - 50 | 123 | 90 (73.2%) | 58 (64.4%) |
| | | 51 - 65 | 79 | 52 (65.8%) | 35 (67.3%) |
| | | above 66 | 96 | 66 (68.8%) | 38 (57.6%) |
| | Sex | Female | 244 | 168 (68.9%) | 104 (61.9%) |
| | | Male | 221 | 143 (64.7%) | 76 (53.1%) |

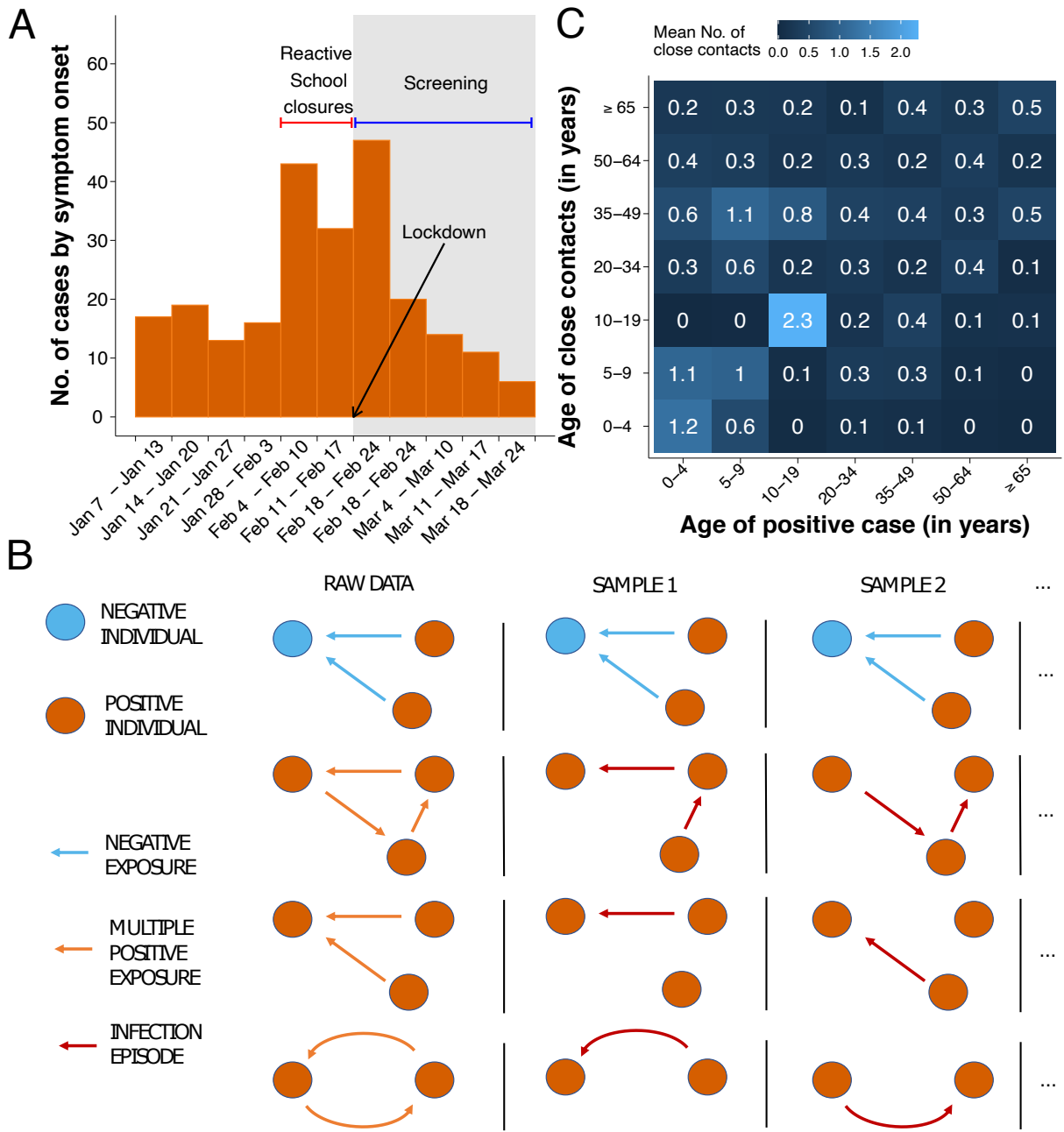

**Figure 1. A)** Time series of symptomatic cases reported in the municipality of Mede by week of symptom onset. **B)** Schematic representation of the sampling algorithm adopted to reconstruct the transmission chains in the sample. **C)** Contact matrix representing the average number of close contacts reported by positive cases.

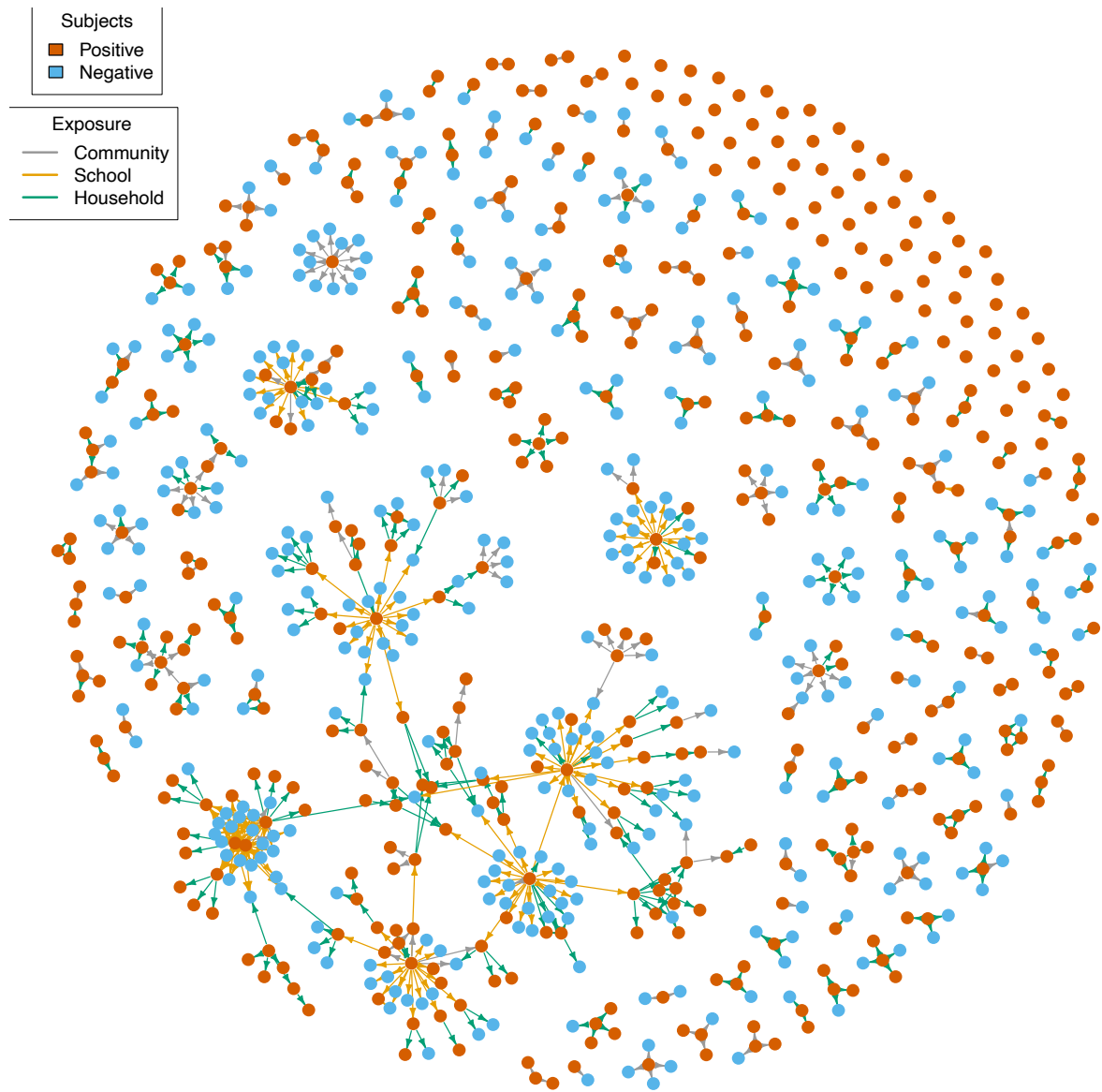

**Figure 2.** Contact networks representing all identified exposure events. The color of the nodes represents the infectious status of each tested individual. Subjects who experienced both negative and positive exposures (namely, 5 individuals) are represented twice. Edges represent the exposure event between two subjects, therefore connecting a positive case to his/her close contacts. The color of the edges represents the setting of exposure.

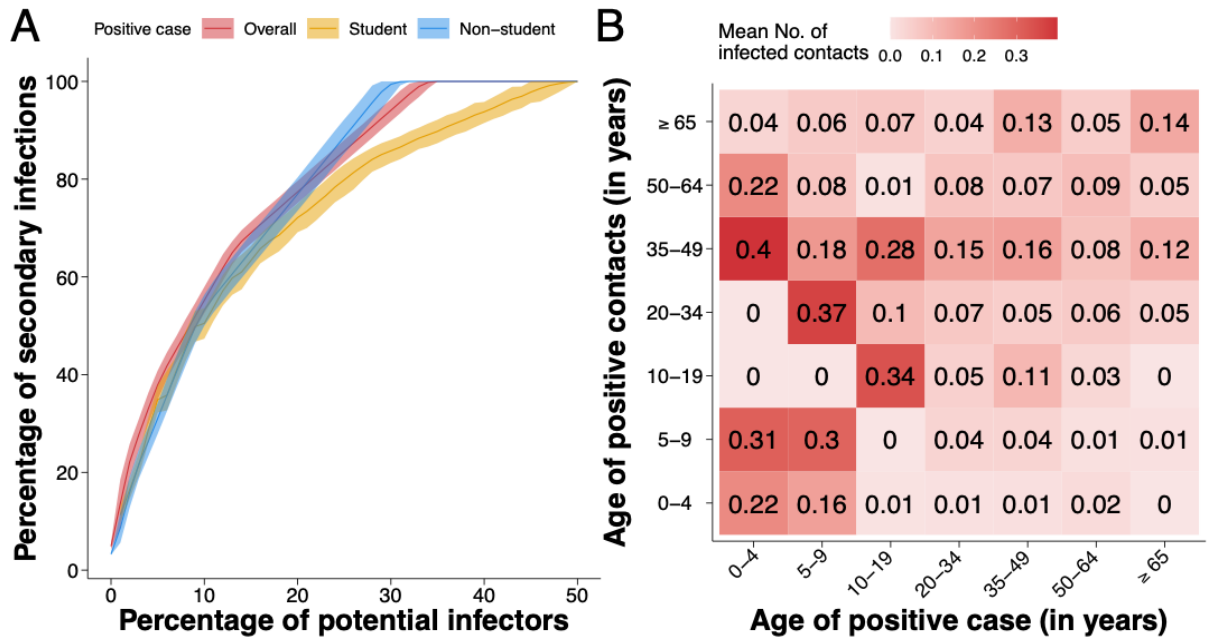

**Figure 3. A)** Distribution of secondary infections generated by identified positive cases. **B)** Transmission matrix representing the average number of infections caused in each age group by positive cases of different ages.